\documentclass[final,1p,times]{elsarticle}

\usepackage{graphicx}
\usepackage{amssymb}
\usepackage{float}
\usepackage{amsmath}

 \biboptions{sort&compress}

\makeatletter
\def\ps@pprintTitle{%
\let\@oddhead\@empty
\let\@evehead\@empty
\def\@oddfoot{}%
\let\@evenfoot\@oddfoot}
\begin{document}

\begin{frontmatter}


\title{Effects of mode mixing and avoided crossings on the transverse spin in a metal-dielectic-metal sphere}



\author{Sudipta Saha$^a$, Ankit Kumar Singh$^a$, Nirmalya Ghosh$^a$, and Subhasish Dutta Gupta$^{b,*}$}
\address[1]{Department of Physical Sciences, Indian Institute of Science Education and Research Kolkata,\\ Mohanpur-741 246, India}
 \address[2]{School of Physics, University of Hyderabad, Hyderabad 500046, India\\  $^*$Corresponding author: sdghyderabad@gmail.com}

\begin{abstract}
We study transverse spin in a sub-wavelength metal-dielectric-metal (MDM) sphere when the MDM sphere exhibits avoided crossing due to hybridization of the surface plasmon with the Mie localized plasmon. We show that the change in the absorptive and dipersive character near the crossing can have significant effect on the transverse spin. An enhancement in the transverse spin is shown to be possible associated with the transparency (suppression of extinction) of the MDM sphere. The effect is attributed to the highly structured field emerging as a consequence of competition of the electric and magnetic modes.
\end{abstract}


\end{frontmatter}


Light-matter interaction in the strong coupling regime resulting in vacuum-field Rabi splittings has been one of the central themes in cavity quantum electrodynamics \cite{Mondragon_1983,Agarwal_1984,Raizen_1989,optcomm1995,vahala2004}. The coupling occurs when the dispersion branches of the uncoupled systems cross and results in the avoided crossing phenomenon and normal mode splitting. The resulting physics can have interesting implications for various applications ranging from fast and slow light \cite{shimizu_2002,manga2004} and optical sensing \cite{hanOL2010} to counting and sizing of nanoparticles \cite{sahin2010}. The avoided crossing phenomena have been observed in a variety of optical as well as condensed matter systems. These include planar or spherical plasmonic or guided wave structures, metamaterial cavities, in photonic crystal fibers \cite{hanOL2010,slowlight_sdg,Rohde_2007,jansen2011}, etc. Contrary to the belief that the split modes can be resolved only with high-finesse optical modes, recent focus on systems with substantial losses (eg. plasmonic nanocavities, and leaky cavities) demonstrated the avoided crossing phenomena \cite{reithmaier2004,benz2013,lien2016} in such systems. Our interest in the strongly coupled systems is motivated by the fact that there is a significant change in the dispersive and absorptive properties near the avoided crossing caused by mode mixing and exchange \cite{manga_thesis, manga2005, pavan2013}. This could result in a highly structured field which is essential for observing yet another important recent discovery, namely, the transverse spin \cite{Belinfante_1940} in optical systems \cite{Aiello_2015,measuring_2015,natphys_2016}.\\
\begin{figure}[ht]
	\centering
	\includegraphics[width=0.6\linewidth,keepaspectratio]{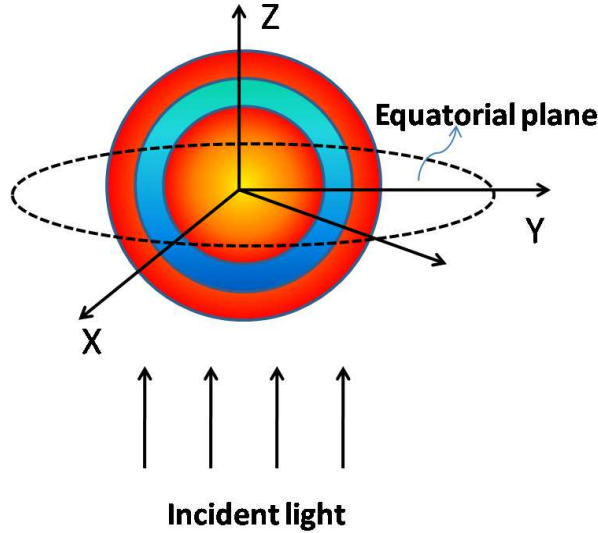}
	\caption{\label{fig1}Schematic view of the scattering geometry for a MDM (Ag-Si-Ag) sphere centered at origin.}
\end{figure}

It is well known that the linear momentum ($\boldsymbol{\mathcal{P}}$) carried by light waves can be decomposed into orbital ($\boldsymbol{\mathcal{P}_o}$) and spin ($\boldsymbol{\mathcal{P}_s}$) parts \cite{allenbook,ngbook}. The orbital momentum ($\boldsymbol{\mathcal{P}_o}$) is the so-called canonical momentum of light which is responsible for the energy transport and radiation pressure. The spin momentum, on the other hand, has long been known as a `virtual' entity which does not transport energy or exerts pressure on the dipolar particle and is only responsible for generating the spin angular momentum (SAM) ($\boldsymbol{\mathcal{P}_s} \approx\nabla\times \mathbf{S}$) \cite{Belinfante_1940}. However, recent studies have shown that such an elusive quantity can also be experimentally observed in case of structured optical fields (e.g. evanescent field, interference of fields, optical vortices etc.) \cite{natphys_2016,Bliokh2014}. This extraordinary transverse spin momentum has been observed to produce a helicity-dependent transverse optical force through higher order (multipolar) interactions with probe particles. Note that usually, SAM (\textbf{S}) has only longitudinal component (along wave vector \textbf{k}) and is associated with the circular/elliptical polarization of light waves with helicity $\sigma $ in the range $-1 \leq\sigma\leq +1$ \cite{ngbook}. In contrary, the transverse SAM has also been observed for structured fields which is independent of helicity \cite{Aiello_2015,measuring_2015,Bliokh2014,Bliokh_2015,spinsurface_2012,Bauer_2016}. These two unusual entities, namely, the polarization-dependent transverse momentum and the polarization-independent transverse SAM observed in evanescent fields (e.g., for surface plasmon-polaritons at dielectric-metal interfaces) have led to several fundamental consequences \cite{Bliokh_2015,Bauer_2016,natphot_2015}. We have recently shown that such rich structure of momentum and spin densities of light can also be observed in scattering of plane waves from micro and nano optical systems \cite{Saha_2016}. We have highlighted the importance of interference and competition of dominant modes of the structure warranting further investigation. Such studies may provide new insights and understanding on these recently discovered non-trivial spin and momentum components of light.\\

\begin{figure}[ht]
	\centering
	\includegraphics[width=0.8\linewidth]{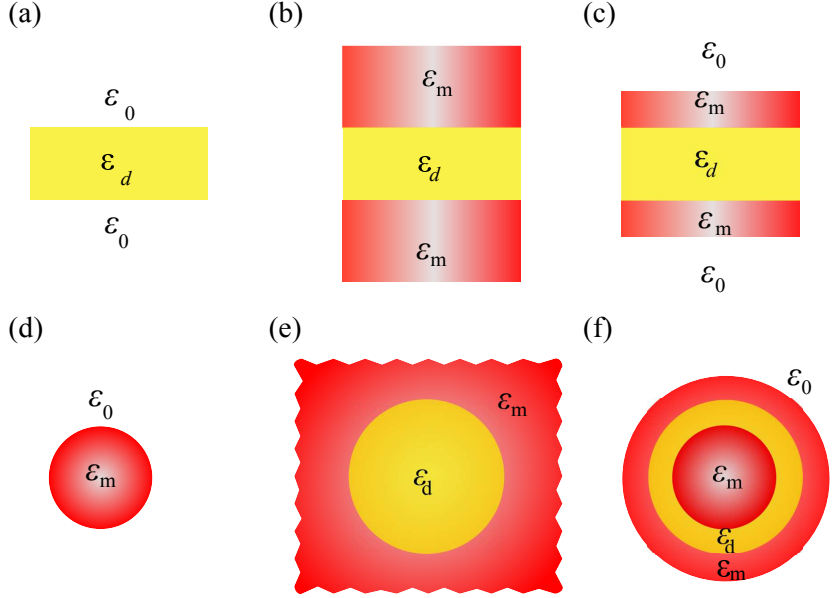}
	\caption{\label{fig2} Analogy between planar and spherical MDM structures. (a) a planar  dielectric guide in air; (b) a planar dielectric guide with semi-infinite metal claddings; (c) planar dielectric guide with thin metal cladding on both sides placed in air. Spherical analog of the planar structure: (d) a single metal sphere in air; (e) a dielectric sphere placed in an infinite metal surrounding; (f) the MDM sphere in air.}
\end{figure}

\begin{figure}[ht]
	\centering
	\includegraphics[width=0.7\linewidth]{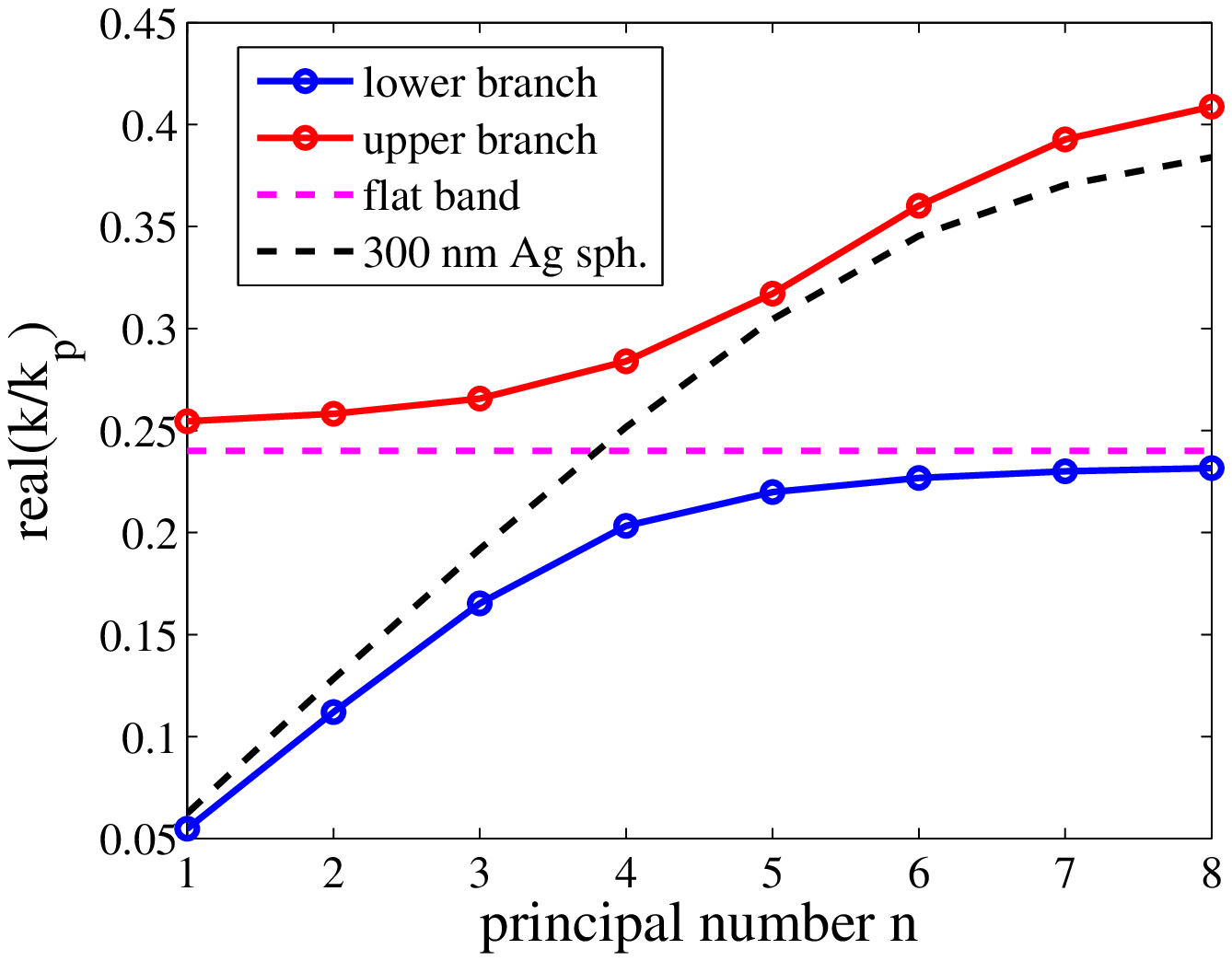}
	\caption{\label{fig3} Dispersion curve: The black dashed line represents the dispersion relation for a Ag sphere of radius R=300 nm placed in air. The upper and lower branch solutions of the dispersion curve for a Ag-Si-Ag sphere placed in air having dimension $R=300 nm$, $L=30 nm$ and $T=10 nm$ are shown by the red and blue solid lines respectively. The flat band of the dispersion curve is shown by the magenta dashed line. }
\end{figure}
In this letter we report the effects of mode coupling in a sub-wavelength metal-dielectric-metal (MDM) sphere on this transverse spin of the scattered waves. For incident circular polarized light, we use full Mie expansions to calculate the electric and magnetic fields inside and outside the MDM sphere to extract the distribution of the Poynting vector $\mathbf{P}$ and the SAM density $\mathbf{S}$. Our discussion starts with the generic features of the components of the Poynting vector and the SAM density and their dependence on the helicity of the incident wave. We next present a brief analysis of the origin of the avoided crossing drawing an analogy with a planar structure \cite{slowlight_sdg,ngbook}. Recall that the dispersion characteristics and the avoided crossing was studied earlier in great detail \cite{Rohde_2007} reporting the spectral tunability of the transparency by changing the width of the dielectric layer. We reproduce similar results albeit for a subwavelength MDM sphere. We further show that this transparency is a consequence of the competition of the electric and magnetic modes. We then look into the spatial distribution of the Poynting vector ($\mathbf{P}$) and SAM density ($\mathbf{S}$) components across the avoided crossing. We show the feasibility of enhancing the transverse SAM density near the avoided crossing. The enhanced transverse SAM density is confirmed by looking at the electric field circulation in the relevant planes near the metal-dielectric boundary. Analogous enhancement in transverse spin has recently been reported in a gap plasmon guide exploiting the mode coupling phenomenon and coherent perfect absorption \cite{sdgEPJAP_2016}. Analogous means of enhancement of the transverse spin is of utmost importance because of the difficulty in measuring the tiny magnitude of the force (order of femto Newton) exerted on a nano particle, which was reported in a recent experiment \cite{natphys_2016}.\\

Consider a layered MDM sphere of total radius $W$ comprising of a metal core with radius $R$, a dielectric layer with thickness  $L$ and a metallic shell with thickness $T$ placed in air (Fig.\ref{fig1}). Let the MDM sphere be illuminated by a plane wave along the positive $Z$-direction. We use full Mie theory \cite{bohren} to calculate the electric $\mathbf{E}$ and magnetic $\mathbf{H}$  fields. The expressions for the $\mathbf{E}$ and $\mathbf{H}$ can be used to calculate the components of SAM density $\mathbf{S}$ and Poynting vector $\mathbf{P}$ defined as  (see, for example, \cite{Berry2009}): \\
\begin{eqnarray}
\mathbf{S}&=&\dfrac{Im(\epsilon \mathbf{E^*} \times \mathbf{E}+\mu \mathbf{H^*} \times \mathbf{H})}{\omega (\epsilon |\mathbf{E}|^2+\mu |\mathbf{H}|^2)} \label{eq1} \\
\mathbf{P}&=&\dfrac{1}{2} Re(\mathbf{E}\times\mathbf{H}^*) \label{eq2}
\end{eqnarray}
Irrespective of the scattering system and its dimension, the following symmetry properties were obtained for input right/left circular polarizations (RCP/LCP) \cite{Saha_2016}:
\begin{eqnarray}
\nonumber &(S_r)_{RCP}=-(S_r)_{LCP};& ~~(P_r)_{RCP}=(P_r)_{LCP} \\
\nonumber &(S_\theta)_{RCP}=-(S_\theta)_{LCP};&~~ (P_\theta)_{RCP}=(P_\theta)_{LCP} \\
&(S_\phi)_{RCP}=(S_\phi)_{LCP};&~~(P_\phi)_{RCP}=-(P_\phi)_{LCP}
\label{eq3}
\end{eqnarray}
Thus $S_r$ and $P_r$ give the conventional longitudinal SAM and momentum, respectively. $S_\theta$ and $P_\theta$ also follow the usual behaviour. The important components are the transverse SAM ($S_\phi$) and transverse Poynting vector ($P_\phi$). The transverse SAM is independent of input helicity, which results from the phase shifted longitudinal field components; whereas, transverse Poynting vector is spin-dependent and it represents polarization-dependent transverse (spin) momentum. \\
\begin{figure}[ht]
	\includegraphics[width=\linewidth]{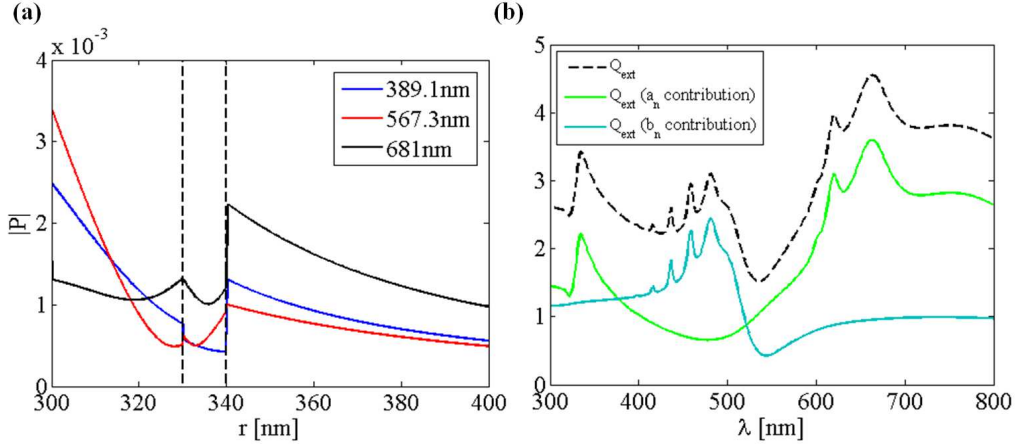}
	\caption{\label{fig4}(a) radial variation of the
		magnitude of $\mathbf{P}$ at wavelength below (blue), at (red) and above (black) the crossing; (b) wavelength variation of $Q_{ext}$ for the same MDM sphere is shown by the black dashed line. The contribution of TM (TE) modes mode is shown by the green (violet) solid curve. The other parameters are as in Fig. \ref{fig3}. The vertical lines represent the inner and outer radii of the metal shell.}
\end{figure}
\par
There has been a great deal of research on plasmon hybridization in metal-dielectric core-shell structures \cite{nordlander_science,Rohde_2007}.
The dominant mechanism of avoided crossing in the MDM sphere under study can be easily understood from a simple analogy to a planar gap plasmon guide \cite{slowlight_sdg}. In a simplified approach to understand the level crossing, the gap plasmon guide of Fig.\ref{fig2}(c) can be thought of as a hybrid of the two limiting structures, namely, a dielectric waveguide (Fig.\ref{fig2}(a)) and a gap plasmon guide with semi-infinite metal extents (Fig.\ref{fig2}(b)).  The dispersion branches of one can cross with the dispersion branches of the other leading to the avoided crossing phenomenon of the gap plasmon guide with finite metal claddings. A direct analogy with the MDM sphere (comparison of the upper and lower panels of Fig.\ref{fig2}) reveals that the avoided crossing in this case results from an interaction of the Mie modes of the metallic core (Fig.\ref{fig2}(d)) with the surface plasmons at the surface of the dielectric sphere in infinite metal surrounding (Fig.\ref{fig2}(e)) (approximated by flat-surface plasmons). Thus the strength of coupling can be controlled by the width of the dielectric  layer. For our simplified model the  crossing of the corresponding dispersion branches  for the structures of Fig.\ref{fig2}(d) and Fig.\ref{fig2}(e) are shown by the black and magenta dashed lines respectively (see Fig.\ref{fig3}). We have plotted the real part of the normalized propagation constant $k/k_p$ ($k_p=2\pi/\omega_p$, $\omega_p$ is the plasma frequency) as a function of the principal mode number $n$. We have also looked at the imaginary part of the propagation constant (not shown). For calculations (for example for  the TM modes) we have used an Ag-Si-Ag sphere with the following set of parameters: $R=300$ nm, $L=30$ nm and $T= 10$ nm.  The flat band in the dispersion curve corresponds to the surface plasmon frequency $k_{sp}/k_p=\omega_{sp}/\omega_p=\frac{1}{\sqrt{\epsilon_b+\epsilon_d}}=0.2401$ where $\epsilon_b=5.1$ (the background dielectric constant in the Drude model for Ag) and $\epsilon_d=12.25$ (for Si) \cite{Rohde_2007}. 
\begin{figure}[ht]
\includegraphics[width=\linewidth,keepaspectratio]{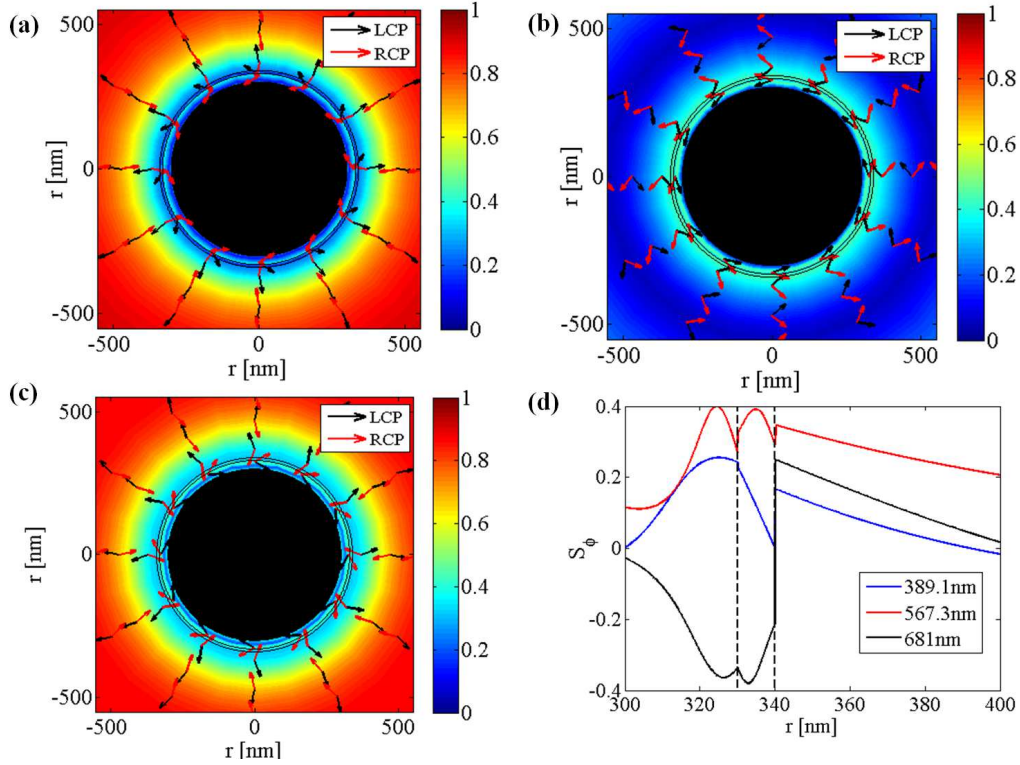}
\caption{\label{fig5} The computed SAM density $\mathbf{S}$ at the equatorial plane of the Ag-Si-Ag sphere for input LCP/RCP is shown at wavelength (a) 389.1 nm (below avoided crossing) (b) 567.3 nm (at the avoided crossing)  and (c) 681nm (above the avoided crossing ); (d) the radial variation of the transverse SAM density ($S_\phi$) at the three wavelengths. In Fig. (a),(b),(c), the inner black solid circle represents the core metal sphere; interface between the middle Si layer and the outer Ag layer, and the outer layer of the MDM sphere are then shown by the black circles outside the core. These interfaces were represented by the black dotted lines in Fig. (d). The other parameters are as in Fig. \ref{fig3}.}
\end{figure}
The upper and lower branch solutions of the  MDM sphere (Fig.\ref{fig2}(f)) placed in air are shown by the red and blue solid curves, respectively.  A  well-defined avoided crossing can be discerned near the principal mode number n=4. Note that in the exact hybridization model \cite{nordlander_science} where the coupling is between the plasmons of the metal core with those of the thin metal shell, the thickness of the metallic shell can also play an important role. Our results on avoided crossings are exact and the simplified model was used just to interpret the avoided crossing phenomenon (the dashed lines in Fig.\ref{fig3}).\\

We begin by looking at some of the features of scattering from the MDM sphere which was noted earlier \cite{Rohde_2007}. It was shown that the avoided crossing phenomenon in the MDM sphere can be used for transparency (suppression of extinction). Analogous results for our case are shown in Figs.\ref{fig4}(a) and \ref{fig4}(b). Fig.\ref{fig4}(a) shows the radial variation of Poynting vector across the avoided crossing for a Ag-Si-Ag sphere with $R=300nm$, $L=30nm$ and $T=10nm$ for input circular polarization at the equatorial plane of the sphere. At the crossing, shown by the red solid curve, the distribution of $|\mathbf{P}|$ outside the MDM sphere is significantly suppressed compared to wavelengths away (blue and black curves) from the crossing. The transparency is confirmed by the wavelength scan of the extinction coefficient (see Fig. \ref{fig4}(b)). As noted earlier
Rodhe \textit{et al.} \cite{Rohde_2007}, this transparency can be tuned over the entire visible wavelength range by changing the thickness of the dielectric layer (we do not show those results here). However, we would like to highlight the physical origin of this transparency as due to competition between the transverse magnetic (TM, $a_n$ modes) and transverse electric (TE, $b_n$ modes) modes. In Fig. \ref{fig4}(b), the extinction crossing section for the same MDM (Ag-Si-Ag) sphere over a broad wavelength range is shown by the black dashed line. The contributions of TM and TE modes to the extinction cross-section are shown by the green and violet curves, respectively, in the same figure. Thus it is clear that the dip in the extinction cross-section comes from the competetion between the TM and TE scattering modes and this important aspect was missed in some of the earlier papers. \\
\begin{figure}[ht]
	\includegraphics[width=\linewidth]{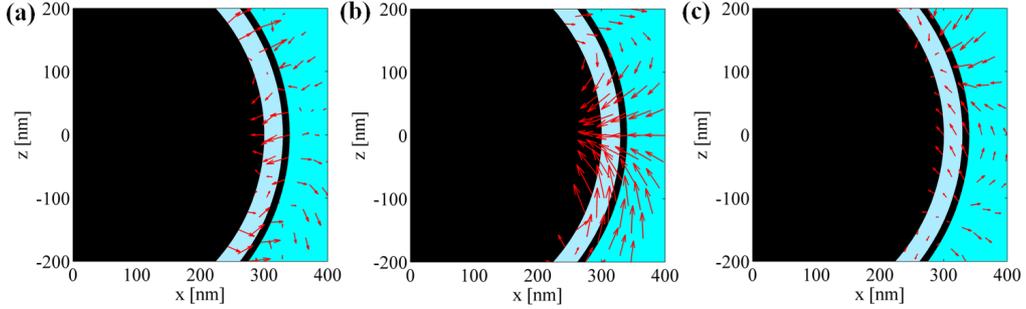}
	\caption{ \label{fig6} The electric field circulation near the metal-dielectric interfaces of the MDM sphere at (a) 389.1 nm (b) 567.3 nm and (c) 681 nm for the MDM sphere with $R=300 nm$, $L=30 nm$ and $T=10 nm$ in the XZ plane. Here the black circle represents the core metal sphere, the internal Si layer is shown by the sky color and the outer metal layer by the black ring. }
\end{figure}

We now present our main results on the transverse spin in the MDM sphere. As noted in the introduction a highly structured field results due to mode hybridization in the MDM sphere. Localization of the fields near the metal-dielectric interfaces is responsible for this structured field which leads to locally enhanced transverse spin. The computed SAM densities $\mathbf{S}$ at the equatorial plane of the Ag-Si-Ag sphere at wavelengths below (389.1nm), at (567.3nm) and above (681nm) the avoided crossing for input LCP/RCP light are shown in Figs. \ref{fig5} (a),(b),(c), respectively. It is clear that the helicity of the beam controls the overall direction of the spin, while the azimuthal component $\mathbf{S}_{\phi}$ is independent of helicity as indicated by Eq.(\ref{eq3}). Moreover, the direction of $\mathbf{S}$ at the crossing (Fig. \ref{fig5}(b)) is observed to be mainly dominated by the azimuthal component and at wavelengths away from the crossing (Fig. \ref{fig5}(a),(c)), the contribution of azimuthal component is significantly reduced. The radial variation of the transverse SAM density $S_\phi$ at the three wavelengths are shown in Fig.\ref{fig5}(d). Thus it is clear that the transverse SAM density gets enhanced for input circular polarization at the avoided crossing of the MDM sphere and as one moves away from the crossing, the transverse SAM density gets reduced. In order to have a clear physical insight, we have studied the circulation of the field in the plane perpendicular to the equatorial plane (i.e. in the XZ plane). In fact this circulation, like in the case of evanescent waves \cite{Bliokh2014}, is the source of the transverse spin. In Fig.\ref{fig6} (a),(b) and (c), the electric field distribution in the XZ plane of the MDM sphere are shown at wavelengths 389.1nm, 567.3nm and 681nm, respectively, near the metal-dielectric interfaces of the MDM sphere. From these figures, it is evident that at the avoided crossing the field becomes highly rotational near the metal-dielectric interfaces (Fig.\ref{fig6}(b)) as compared to wavelengths away from crossing (Fig.\ref{fig6}(a),(c)). This clearly explains the origin of the enhanced transverse SAM density at the crossing for input circular polarization. Another important aspect of the transverse spin in the MDM sphere needs to be stressed. In planar structures supporting surface plasmons the only contribution to the extraordinary spin comes from the $\mathbf{E}^*\times \mathbf{E}$ with null contribution from the magnetic counterpart, since $\mathbf{H}$ has only one nonvanishing component, perpendicular to the plane of incidence. In the spherical counterpart there can be contributions from both the electric and the magnetic fields (see Eq. (\ref{eq1})), with varying strengths below and above the crossing. For example, below the crossing, the contribution from the magnetic field part can be significant, while at and above it the same can be nominal (results not shown). \\

In summary, we have studied the effect of the avoided crossing on the SAM density and Poynting vector components in a MDM sphere. The results show that the transverse SAM density is enhanced at the crossing while the Poynting vector gets suppressed outside the MDM structure resulting in the observed transparency. These results are supported by explicit quiver plots of the electric field near and away from the crossing. We further show that the observed optical transparency originates from a competition between transverse electric and magnetic modes.\\

Authors would like to thank Shourya Dutta-Gupta for many helpful discussions. SS thanks Indian Institute of Science Education and Research Kolkata (IISER-K) for the research fellowship and the School of Physics, University of Hyderabad (UoH) for providing local hospitality during his stay at the UoH campus. AKS acknowledges the Council of Scientific and Industrial Research (CSIR) for the research fellowship.





\bibliographystyle{model1-num-names}
\bibliography{mdmref.bib}







\end{document}